\documentclass[prl, reprint, superscriptaddress]{revtex4-1}
\usepackage{graphicx}
\usepackage{color}
\usepackage{amsmath}
\usepackage{CJK}

\begin{document}
\begin{CJK*}{UTF8}{mj}

\title{Probing strongly hybridized nuclear-electronic states in a model quantum ferromagnet}

\author{I. Kovacevic}
\email{ivankowacevic@gmail.com}	
\affiliation{Laboratory for Quantum Magnetism, Institute of Physics, Ecole Polytechnique F\'{e}derale de Lausanne (EPFL), CH-1015 Lausanne, Switzerland}
\author{P. Babkevich}
\email{peter.babkevich@gmail.com}	
\affiliation{Laboratory for Quantum Magnetism, Institute of Physics, Ecole Polytechnique F\'{e}derale de Lausanne (EPFL), CH-1015 Lausanne, Switzerland}
%\author{M. Jeong}
\author{M. Jeong (정민기)}
\email{minki.jeong@gmail.com}	
\affiliation{Laboratory for Quantum Magnetism, Institute of Physics, Ecole Polytechnique F\'{e}derale de Lausanne (EPFL), CH-1015 Lausanne, Switzerland}
\author{J. O. Piatek}
\affiliation{Laboratory for Quantum Magnetism, Institute of Physics, Ecole Polytechnique F\'{e}derale de Lausanne (EPFL), CH-1015 Lausanne, Switzerland}
\author{G. Boero}
\affiliation{Microsystems Laboratory, Ecole Polytechnique F\'{e}derale de Lausanne (EPFL), CH-1015 Lausanne, Switzerland}
\author{H. M. R\o nnow}
\affiliation{Laboratory for Quantum Magnetism, Institute of Physics, Ecole Polytechnique F\'{e}derale de Lausanne (EPFL), CH-1015 Lausanne, Switzerland}

\begin{abstract}
We present direct local-probe evidence for strongly hybridized nuclear-electronic spin states of an Ising ferromagnet LiHoF$_4$ in a transverse magnetic field. The nuclear-electronic states are addressed via a magnetic resonance in the GHz frequency range using coplanar resonators and a vector network analyzer. The magnetic resonance spectrum is successfully traced over the entire field-temperature phase diagram, which is remarkably well reproduced by mean-field calculations. Our method can be directly applied to a broad class of materials containing rare-earth ions for probing the substantially mixed nature of the nuclear and electronic moments.
\end{abstract}

\maketitle
\end{CJK*}

\begin{figure}
\centering
\includegraphics[width=0.45\textwidth]{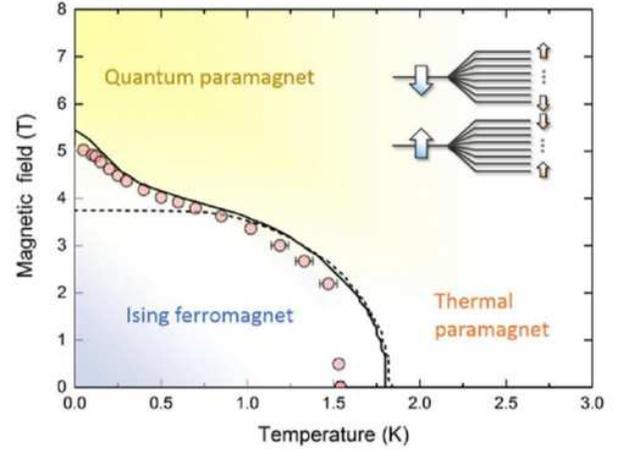}
\caption{Phase diagram of LiHoF$_4$ in a transverse magnetic field with the experimental data taken from Ref.~\cite{Bitko96PRL}. Solid line represents a mean-field calculation following Ref.~\cite{Ronnow07PRB} taking into account strong hyperfine interaction, while dashed line is calculated without hyperfine interaction. Inset shows schematic energy levels for the Ising spins in the ordered phase (left) and its modification by hyperfine interactions with the nuclear spins (right). 
\label{fig:phase}}
\end{figure}

The compound $\mathrm{LiHoF_4}$ is widely regarded as a prototypical system realizing the transverse-field Ising model \cite{Sachdev}. The groundstate in zero field is ferromagnetically ordered, while applying a relatively small transverse field induces a zero-temperature quantum phase transition at $H_c=4.95$ T into a quantum paramagnet \cite{Bitko96PRL}, as shown in Fig.~\ref{fig:phase}. Meanwhile, the hyperfine coupling strength of a Ho$^{3+}$ ion is exceptionally large with a coupling constant $A=39(1)$ mK \cite{Magarino80PRB, Mennenga84JMMM}. The resulting strong hybridization between the electronic and nuclear magnetic moments~\cite{hybrid} leads to two dramatic effects close to the quantum critical point: (i) significant modification of the low-temperature magnetic phase boundary (see Fig.~\ref{fig:phase})~\cite{Bitko96PRL}; (ii) incomplete mode softening of the low energy electronic excitations at the critical point~\cite{Ronnow05Sci}. Therefore, this system  provides a rare opportunity to explore the quantum phase transition of a magnet coupled to a nuclear spin bath \cite{Bitko96PRL, Ronnow05Sci, Ronnow07PRB, Babkevich15PRB}.

The impact of strong hybridization has also been highlighted for magnetic-ion diluted insulators, such as LiYF$_4$:Ho$^{3+}$ using magnetic resonance~\cite{Giraud01PRL, Giraud03PRL}. A similar line of effort has achieved more recently single-molecule magnetic resonance with a rare-earth ion~\cite{Mullegger14PRL}. Furthermore, strong hybridization is of great interest in quantum information science~\cite{Morley10NatM, Morley13NatM, Shiddiq16Nat}. As much as these examples focus on the single-ion limit, the other limiting case of many-body systems, such as LiHoF$_4$, provides a very different and complementary perspective. While in the long-range-ordered state the hybridization is suppressed, an applied transverse field introduces quantum fluctuations enhancing the hybridization towards $H_c$.

However, probing directly the strongly hybridized states in LiHoF$_4$ using spectroscopic methods, at the lowest energy scale, has so far been restricted to the thermal paramagnetic phase in the single-ion limit. The involved energy scale is too low to be resolved by the neutron scattering \cite{Ronnow05Sci, Ronnow07PRB}. Magnetic resonance on $^{165}$Ho nuclei would provide a direct way of probing the hybridized nuclear-electronic states. However, the resonance in the ordered phase is expected around the frequency of 4.5 GHz in zero field, which does not fall into the operating frequencies of conventional nuclear magnetic resonance (NMR) nor electron spin resonance (ESR) instrumentation. Some studies have reported a hyperfine structure in ESR \cite{Magarino80PRB, Janssen85JPCS}, but all in the paramagnetic regime above the ordering temperature $T_c=1.53$ K \cite{Bitko96PRL}. To date, microscopic evidence for the realization of the unique nuclear-electronic Ising model \cite{Schechter05PRL, Schechter08PRB} is absent.

Here we demonstrate experimentally nuclear-electronic magnetic resonance in LiHoF$_4$ using coplanar microwave resonators and a vector network analyzer (VNA). We successfully trace the temperature and field evolution of the spectrum over the entire phase diagram, and show that it is remarkably well reproduced by a mean-field calculation with parameters set by independent spectroscopic measurements \cite{Babkevich15PRB, Magarino80PRB, Mennenga84JMMM, Ronnow07PRB}.

\begin{figure}[!t]
\centering
\includegraphics[width=0.5\textwidth]{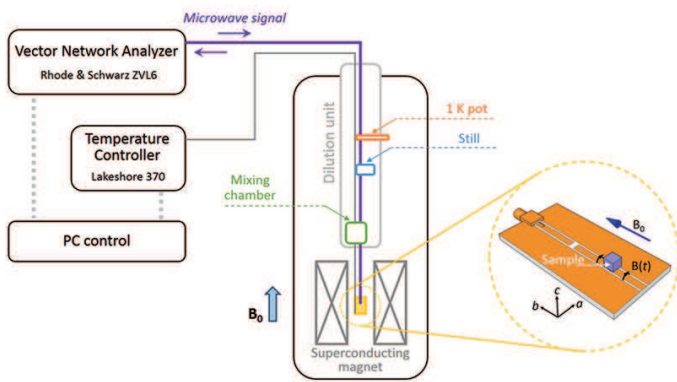}
\caption{Schematic drawing of the setup showing the sample mounted on top of the microwave coplanar resonator inside the vertical field magnet.
\label{fig:setup}}
\end{figure}

We begin with a description of our experimental setup shown in Fig.~\ref{fig:setup}. A series of microwave coplanar resonators with different fundamental frequencies from 1.7 to 5.6 GHz were prepared. The impedance of the resonator is matched to the rest of the system by optimizing the gap size between the conductors. The oscillating magnetic field, {\bf B}({\it t}), generated at the sample position is parallel to the surface. A cube shaped sample of $2\times2\times2$ mm$^3$ was placed at the center of the active strip, with a sub-millimeter gap in-between to avoid unwanted heating. The measurement geometry was chosen such that the applied magnetic field, {\bf B}$_0$, is along the crystallographic {\it b} axis of the tetragonal Scheelite structure, and {\bf B}({\it t}) is perpendicular to both {\bf B}$_0$ and the {\it c} axis to satisfy the magnetic resonance condition. We measured the $S_{11}$ parameter, which is defined as the ratio of the reflected to the input power, using a VNA which is connected through a low-loss cryogenic coaxial cable to the coplanar resonator. The coaxial cable was thermally anchored at each stage of the dilution refrigerator including the 1 K pot, Still, and mixing chamber to ensure thermalisation. The sample thermometer was located only 5 mm away from the sample which was thermally anchored to the mixing chamber. With an input power of -16 dBm applied by the network analyzer, the sample base temperature was 0.15 K to within 0.01 K.

\begin{figure*}[!t]
\centering
\includegraphics[width=1\textwidth]{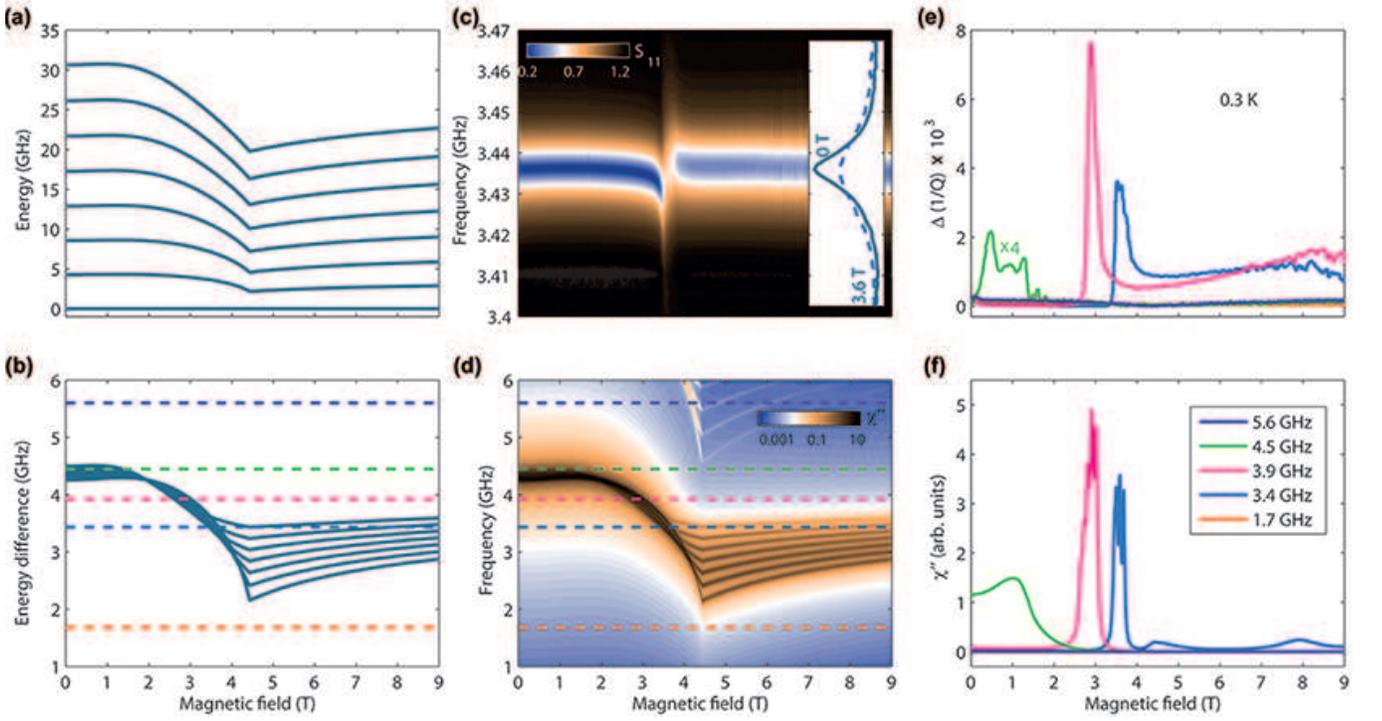}
\caption{(a) Calculated energy levels for the nuclear-electronic groundstate in the mean-field approximation as a function of transverse magnetic field, and (b) the field evolution of the corresponding energy difference between the adjacent levels. The calculations were performed for 0.3 K to compare with the experiments. Magnetic resonance would occur when the excitation frequency (dashed lines) intersects with the energy difference. (c) Frequency-field map of the experimental $S_{11}$ parameter, where the anomaly around 3.6 T corresponds to the expected resonance field for the frequency of 3.4 GHz. Inset shows constant-field cuts of the $S_{11}$ map at (dashed) and away (solid) from the resonance. (d) Frequency-field map of the calculated $\chi''$ intensity. (e) Experimental magnetic resonance spectra obtained for several different frequencies, and (f) the calculated $\chi''$ for the corresponding frequencies.
\label{fig:freq}}
\end{figure*}

To guide and interpret our experimental investigation, we perform a model calculation using a mean-field approximation. The full Hamiltonian has been well characterized through a number of different experiments \cite{Magarino80PRB, Ronnow07PRB} and is given by,
\begin{eqnarray}
\mathcal{H} &=&\sum_i \left[\mathcal{H}_{\rm CF}(\mathbf{\hat{J}}_i) + A \mathbf{\hat{J}}_i \cdot \mathbf{\hat{I}}_i- g_L\mu_{\rm B} \mathbf{\hat{J}}_i\cdot \mathbf{B_0} \right] \nonumber \\
& & -\frac{1}{2}\sum_{ij} \sum_{\alpha\beta}\mathcal{J}_{\rm D} {{D}}_{\alpha\beta} \hat{J}_{i\alpha} \hat{J}_{j\beta} - \frac{1}{2}\sum_{\langle ij \rangle}\mathcal{J}_{\rm ex}\mathbf{\hat{J}}_i \cdot \mathbf{\hat{J}}_j \,
\label{eq:eq1}
\end{eqnarray}
where $\mathbf{\hat{J}}_i$ ($J = 8$) and $\mathbf{\hat{I}}_i$ ($I = 7/2$) are the electronic and nuclear angular momentum operators at site $i$, the dipolar coupling constant $\mathcal{J}_{\rm D} = n (g_L \mu_B)^2 = 13.5$ mK, $D_{\alpha\beta}$ is the dimensionless coupling parameter for the dipole-dipole interaction \cite{Jensen91}, and the negligible exchange constant $J_{\rm ex} = -1.2$ mK. The nuclear Zeeman and quadrupole interactions are assumed to be negligible~\cite{Giraud01PRL}. The crystal field interaction $\mathcal{H}_{\rm CF}$ with the surrounding ions splits the electronic states resulting in a groundstate which is a non-Kramers doublet with a strong Ising-like anisotropy and the first excited state 11\,K above. In the ordered state, dipolar coupling lifts the groundstate degeneracy resulting in pseudo-spins up and down which we label as $\mid\uparrow\rangle$ and $\mid\downarrow\rangle$ states. Each state is further split into 8 nuclear-electronic states by the hyperfine interaction (Fig.~\ref{fig:phase}(a), inset).

The total Hamiltonian can be diagonalized in the basis of $(2J+1)\times(2I+1)=136$ nuclear-electronic $|\alpha \rangle = |m_J, m_I\rangle$ states. The evolution of the lowest states with the applied transverse field is shown in Fig.~\ref{fig:freq}(a).  The energy level difference between consecutive states, $\Delta E$, changes dramatically with the field as illustrated in Fig.~\ref{fig:freq}(b). In the first approximation $\Delta E$ is proportional to $A |\langle \mathbf{J} \rangle|$, where $|\langle \mathbf{J} \rangle|$ is the magnitude of the total angular momentum, hence $\Delta E$ decreases with the field and reaches a minimum at $H_c$. The diagram shown in Fig.~\ref{fig:freq}(b) allows us to predict at which field the magnetic resonance occurs for a given frequency.

Experimentally we observed magnetic transitions between the adjacent nuclear-electronic levels through resonant absorption of continuous microwaves by the sample on a coplanar resonator. Figure~\ref{fig:freq}(c) presents a typical frequency-field map at 0.3 K of the $S_{11}$ parameter using a resonator with the unloaded frequency of 3.4 GHz. The map shows a clear anomaly around 3.6 T indicative of magnetic resonance. This field value indeed agrees with the one predicted by mean-field calculations, which can be seen in Fig.~\ref{fig:freq}(b) by taking an intersect of blue dashed line for 3.4 GHz with the solids lines for the energy level difference.

For an in-depth comparison between experiments and calculations, we proceed to directly calculate the imaginary part of the frequency-dependent susceptibility $\chi''(f)$ which is responsible for magnetic resonance absorption \cite{Cowan05, Abragam82}. The calculations were performed within the linear-response framework \cite{Jensen91} using the mean-field wavefunctions $|\alpha \rangle$ and $|\alpha' \rangle$,
\begin{widetext}
\begin{equation}
\chi'' _{}(f)  =  \sum_{\alpha \alpha'} \frac{
 \langle \alpha | (g_L \mu_B \hat{J}_y + g_N \mu_N \hat{I}_y) | \alpha' \rangle
 \langle \alpha' | (g_L \mu_B \hat{J}_y + g_N \mu_N \hat{I}_y) | \alpha \rangle
}{
(E_{\alpha'} - E_\alpha - hf)^2 + \Gamma^2_{\alpha' \alpha}
}
\Gamma_{\alpha' \alpha} (n_\alpha - n_{\alpha'}) + \chi'_{}(0) ,
\label{eq:eq2}
\end{equation}
\end{widetext}
where $E_\alpha$ is the energy of the hybridized nuclear-electronic eigenstates in the presence of the mean-field, $n_\alpha=\exp(-\beta E_\alpha)/Z$ is the thermal population factor and $Z=\sum_{\alpha'} \exp(-\beta E_{\alpha'})$ is the partition function. The subscript $y$ refers to the oscillating field direction. The lifetime in the linear-response calculation of the states is assumed to be independent of field and temperature, and was fixed to 40 ns, corresponding to a damping of $\Gamma_{\alpha' \alpha} = 0.17$ GHz, which provided the best match to our data. The lifetime broadening may result from direct or indirect contributions from the electronic dipolar and exchange or nuclear dipolar couplings~\cite{Abragam82}, which we leave for future study. We note that the contribution to susceptibility from electronic moments, $\hat{J}_y$, is 500 times larger than the contribution from nuclear moments $\hat{I}_y$. Therefore, despite the predominantly nuclear-spin nature of the $\mid\uparrow\rangle$ levels, the response we measure comes mainly from the electrons. This gives a tremendous enhancement of the signal from the nuclear states amplified by electronic moments. Figure~\ref{fig:freq}(d) presents the calculated frequency-field map of $\chi''$ intensity at 0.3 K, which shows a drastic change upon approaching $H_c$ from below. Resonant absorption is expected from our calculations to be in the 2 to 4.5 GHz bandwidth.

The absorptive part of the susceptibility is experimentally estimated as $\chi'' \propto \Delta(1/Q)$ ~\cite{Cowan05}, where the quality factor $Q$ is defined as the loaded frequency divided by the full-width-half-maximum in the absorption profile in frequency as shown in the inset of Fig.~\ref{fig:freq}(c). In Fig.~\ref{fig:freq}(e) we show the experimental magnetic resonance spectra at 0.3 K for several different frequencies by plotting $\Delta(1/Q) = 1/Q - b$, where $b$ is a uniform background, which can be compared to the calculated spectra at 0.3 K in Fig.~\ref{fig:freq}(f). Both calculations and measurements at frequencies of 3.4 and 3.9 GHz show resonant peaks around 3.6 and 3.0 T, respectively. Conversely, no resonance features are visible for the frequency of 1.7 GHz in both calculations and experiments. The predicted transitions between second-nearest neighbouring levels at 5.6 GHz is too weak to be observed experimentally. The calculated spectrum for 4.5 GHz appears as a broad hump at fields below 2 T, which can be expected from Fig.~\ref{fig:freq}(d) where the frequency line cuts along the strongest $\chi''$ intensity. For a better comparison the $A$ value was slightly reduced by 3\%, which is nearly within the uncertainty from the reported one \cite{Magarino80PRB}. In principle, the uncertainty in the crystal field parameters can influence our calculations \cite{Babkevich15PRB}. Nevertheless, excellent agreement with the experiments is remarkable considering that the model is essentially parameter-free. Some minor discrepancies such as the fine structure in the 4.5 GHz experimental spectrum are likely due to fixed lifetime of all levels in our model. However, since the modes around 4.5 GHz lie very close in the relevant field range, that structure would depend critically on the tiny variations of parameters. We therefore consider it more prudent to use a constant damping. The high-field tails in 3.4 and 3.9 GHz spectra are possibly due to the neglected effects of fluctuations.

\begin{figure}[!h]
\centering
\includegraphics[width=0.35\textwidth]{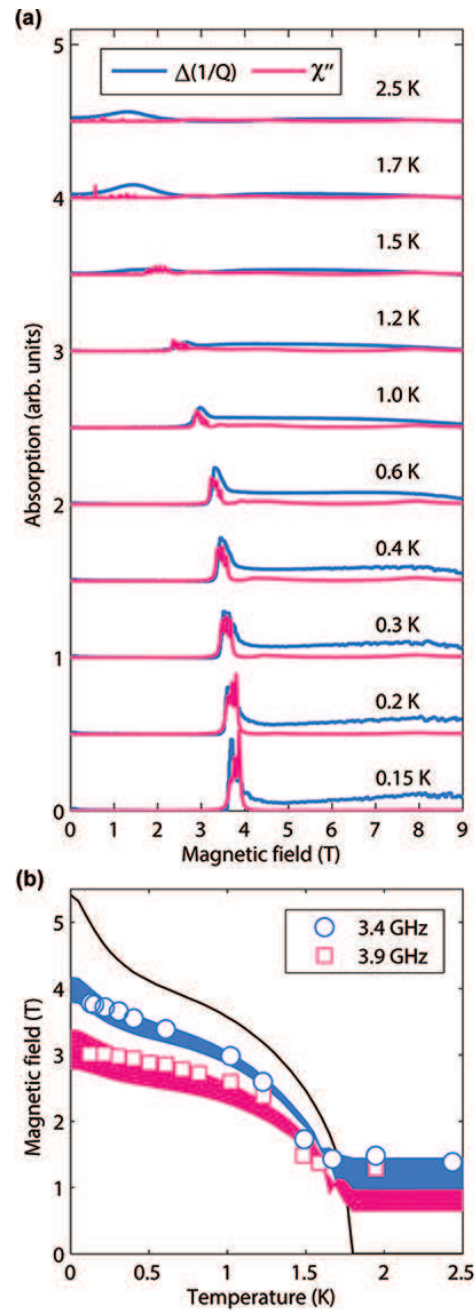}
\caption{(a) Temperature evolution of the spectra from 0.15 K to 2.5 K, from experiments (blue) and calculations (red), using excitation frequency of 3.4 GHz. (b) The resonance field as a function of temperature for two different frequencies. The colored bands are calculations using hyperfine constant in the range of $\pm3$ \% from the value used in Fig.~\ref{fig:freq}(f). Black line reproduces the calculated phase boundary.
\label{fig:temp}}
\end{figure}

Furthermore, we investigate the temperature evolution of the spectrum for 3.4 GHz from 0.15 to 2.5 K as shown in Fig.~\ref{fig:temp}(a). At base temperature a resonance peak appears around 3.7 T, which on warming decreases in amplitude and shifts to lower fields. The former is due to redistribution of the thermal population of states at higher temperatures. The latter reflects the decreasing size of the ordered electronic moment with increasing temperature, sensed by the nuclei through the hyperfine interactions. In Fig.~\ref{fig:temp}(b) we track the resonance field as a function of temperature. Our measurements are shown to be very sensitive to small variations of the hyperfine coupling as depicted by the bands.

As shown in Fig.~\ref{fig:freq} and~\ref{fig:temp}, all the salient features of the experimental results are well reproduced by the model calculations, thereby validating the transverse-field nuclear-electronic Ising model \cite{Schechter05PRL, Schechter08PRB}. The excellent description of the experimental results by our model implies that the probed states have a strongly hybridized character of both nuclear and electronic degrees of freedom. While this has been only hinted by previous bulk measurements \cite{Bitko96PRL} and neutron spectroscopy \cite{Ronnow05Sci}, here we show directly the transitions between the strongly hybridized nuclear-electronic states. Likewise, the presented magnetic resonance should be distinguished from conventional NMR and ESR where the electronic and nuclear moments are approximated to product states \cite{Abragam82, Abragam12, Cowan05}.

\begin{figure}
\centering
\includegraphics[width=0.35\textwidth]{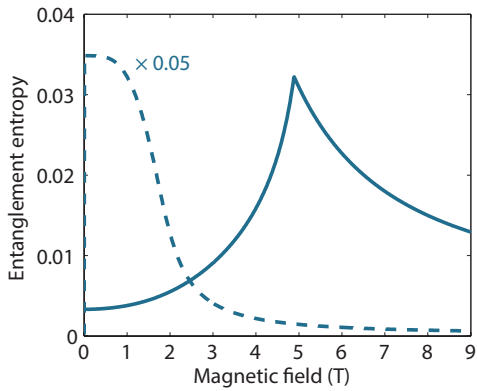}
\caption{Entanglement entropy calculated for the groundstate of LiHoF$_4$ as a function of transverse magnetic field (solid line). Dashed line is the calculation without dipolar interactions. 
\label{fig:entanglement}}
\end{figure}

To highlight qualitative difference in the hybridized states between those in the many-body system and in the single-ion limit, we calculate the groundstate entanglement entropy \cite{Nielsen, Bennett} between the electronic and nuclear moments as a measure of the hybridization. We employ the Schmidt decomposition of the mean-field wavefunction, $|\Psi\rangle = \sum_n c_n |m_J\rangle \otimes |m_I\rangle$, where $c_n \geq 0$ and $\sum_n c^2_n = 1$, where the entanglement entropy is given by the von Neumann entropy $S = - \sum_n |c_n|^2 \ln |c_n|^2 $. The calculated entropy in the absence of dipolar interactions decreases smoothly with a transverse field (Fig.~\ref{fig:entanglement}) in agreement with those reported by Ref. \cite{Schechter08PRB}. However, by turning on dipolar coupling the model produces a cusp-like peak at $H_c$, that is, the hybridization in the ordered state of LiHoF$_4$ increases with the applied field until it reaches a peak at the critical point. The field essentially mixes the higher excited states into the groundstate, thereby enhancing the hybridization. Increasingly larger field, $H>H_c$, magnetizes the electronic and nuclear moments along the field direction such that the groundstate approaches a product state.

To summarize, we have demonstrated Ho nuclear-electronic magnetic resonance of LiHoF$_4$ in a transverse magnetic field over the entire field-temperature phase diagram. The spectral evolution is remarkably well reproduced by mean-field calculations, validating the transverse-field nuclear-electronic Ising model. Taking advantage of the well-characterized model nature of LiHoF$_4$, we have successfully probed the strongly hybridized states and their evolution in the long-range-ordered state. Our experimental scheme will find direct applications not only in the Li$R$F$_4$ ($R$=rare earth) family \cite{Kraemer12Sci, Babkevich15PRB, Babkevich16PRL}, but also other $R$ containing compounds including spin glass \cite{Schechter05PRL, Schechter08PRB, Silevitch07Nat, Ancona-Torres08PRL, Piatek14PRB} and spin ice \cite{Harris97PRL, Sala12PRL}.

We are grateful to M. Graf, S. S. Kim and P. Jorba Cabre for their contribution in building experimental setup at initial stage, B. Dalla Piazza for sharing his insight into the mean-field and linear-response theory. We also thank J. Jensen, A. Feofanov and D. Yoon for helpful discussions.  M.J. is grateful to support by European Commission through Marie Sk{\l}odowska-Curie Action COFUND (EPFL Fellows). This work was supported by the Swiss National Science Foundation, the MPBH network and European Research Council grant CONQUEST. I.K., P.B., and M.J. contributed equally to this work.

\bibliography{bib_LiHoF4}

\begin{thebibliography}{30}
\bibitem [{hyb()}]{hybrid}%
  \BibitemOpen
  \href@noop {} {\bibinfo  {journal} {One may consider the hybridization strong
  once the hyperfine coupling strength is comparable to the Zeeman energy
  scale, for instance, see Ref.~\cite{Morley13NatM}}\ }\BibitemShut {NoStop}%
\end{thebibliography}

\end{document}